\documentstyle[12pt]{article}
\begin{document}

\title{\large \bf Calculation of Chiral
 Lagrangian Coefficients }
\author{ WANG Zhi-Gang $~~~~~~~~~~$ \footnote{E-mail,wangzg@mail.ihep.ac.cn}\\
\small  Institute of High Energy Physics, Chinese Academy of Sciences, Beijing 100039 }
\maketitle

 We present a systematic way to combine the global
 color model and the instanton liquid model to calculate the chiral
 Lagrangian coefficients.  Our numerical results are in agreement well with the
 experimental values.


Quantum chromodynamics (QCD) is successful to describe
strong interaction in terms of basic force at high energy.
However, the large coupling constant at low energy makes the
perturbation theory break down.  At low-energy scale,
the freedoms of quarks and gluons realize themselves in hadronic
level as mesons and baryons, which is characterized
by chiral perturbation theory ($\chi$PT).$^{\cite{Gasser,Eaker}}$
The $\chi$PT can
give a lot of satisfactory explanations for experimental data
while the coefficients of the chiral expansion are input parameters
and their values are determined from experimental observables.
Since the present technique cannot derivate the  chiral
 Lagrangian  from QCD directly,  a class of chiral invariant
 QCD based models are used in deriving the chiral Lagrangian which is
 characterized by the effective quark-quark interaction.

 The global colour model (GCM), which has provided a lot of successful
 descriptions of the long distance properties of strong interaction
 and the QCD  vacuum as well as hadronic phenomena at low energy,
 is based on  a truncation of QCD Lagragian preserves global
 colour gauge  transformation and permits a large $N_{c}$
  expansion.$^{\cite{Cahill,Frank,Roberts,Tandy}}$
   The instanton liquid model (ILM) of the QCD vacuum is based on a
semiclassical approximation, in which all gauge configurations are
replaced by an ensemble of topological nontrivial fields,
instantons and anti-instantons.$^{\cite{Schafer,Hutter}}$
 Although it does not
give rise to a long
range confining force between quarks, the instanton vacuum has been shown
to provide a good phenomenological description of many hadronic properties.
Therefore it is interesting to calculate the chiral coefficients with the
 two coupling nonperturbative models. In fact, the method of  the
 two coupling nonperturbative models to calculate  physical
 quantities already exists in the literature and is
 successful in describing some hadronic phenomena.$^{\cite{Zong}}$
 In this Letter, we approximate
 the effective  quark-quark interaction  by a dilute instanton
 liquid ensemble and calculate the  chiral coefficients from the GCM
 following Ref. \cite{Frank}, while the pioneer work for obtaining
 chiral hadronic phenomena  through
GCM was given by Roberts {\it et al.}$^{\cite{CDRoberts}}$

Here we describe briefly the main skeleton of the GCM.
The global colour model, based upon an effective quark-quark interaction which
is approximated by the dilute instanton liquid configurations,
is defined by a truncation of QCD Lagrangian while maintaining
all global symmetries.

The generation function for QCD in Euclidean space can be written as
\begin{equation}
Z[\bar{\eta}, \eta] = \int {\cal{D}}\bar{q} {\cal{D}} q
{\cal{D}} A \exp\left\{-S
+ \int d^4 x (\bar{\eta} q + \bar{q} \eta)\right\} ,
\end{equation}
where
$$ S = \int d ^4 x \left\{ \bar{q} \left[ \gamma _{\mu}
\left(\partial_{\mu}
- i g \frac{\lambda ^{a}}{2} A^{a}_{\mu} \right)+\psi \right] q +
\frac{1}{4}
G_{\mu \nu}^{a} G_{\mu \nu}^{a} \right\}, $$
and $G_{\mu \nu}^{a}=\partial_{\mu}A^{a}_{\nu}-\partial_{\nu}
A^{a}_{\mu}+gf^{abc}A^{b}_{\mu}A^{c}_{\nu}$.
We leave the gauge fixing term, the ghost field term and its
integration measure to be understood.
Here $A^{a}_{\mu}$ are approximated by
the fields of the dilute instanton liquid.
In order to preserve the chiral invariance
of the above action $S$ (not the full action), the quarks have to be
coupled to a external
field $\psi(x)$ in stead of current quark mass $m$, the field $\psi(x)$
transforms in a certain way under
chiral transformations$^{\cite{Gasser,Eaker,Frank}}$ (For detailed
discussion one can see Ref.\cite{Frank}).

Here we introduce
\begin{equation}
  e^{W[J]}=\int{\cal D}Ae^{\int d^{4}x(-{1\over
  4}G^a_{\mu\nu}G^a_{\mu\nu}+J^a_{\mu}A^a_{\mu})},
\end{equation}
with $J^a_{\mu}(x)=\bar{q}(x)\gamma_{\mu}\frac{\lambda^a}{2}q(x)$ and
rewrite the generation function  as
\begin{equation}
  Z[\psi, \bar{\eta}, \eta]=\int{\cal D}q{\cal D}\bar{q}e^{-\int d^{4}x
  [\bar{q}({\gamma \cdot \partial}+\psi)q -\bar{\eta}q-\bar{q}\eta]}
          e^{W[J]},
\end{equation}
then expand  $W[J]$ in the current $J^a_{\mu}$:
\begin{equation}
W[J]={1\over 2}\int d^{4}xd^{4}y J^a_{\mu}(x)D^{ab}_{\mu\nu}
(x,y)J^b_{\nu}(y)
+\frac{1}{3!}\int J^a_{\mu}J^b_{\nu}J^c_{\rho}D^{abc}_{\mu\nu\rho}+\cdots.
\end{equation}
The GCM is defined by a truncation of the functional $W[J]$ in which
the higher order $n(\ge3)$-point functions are neglected, and only the
gluon 2-point function $D^{ab}_{\mu\nu}(x,y)$ is retained.
This model maintains all global
symmetries of QCD and permits a $1/N_c$ expansion, however
local SU(3) gauge invariance is lost by the truncation. For
simplicity, we use a Feynman-like gauge
$D^{ab}_{\mu\nu}(x-y)=\delta_{\mu\nu}\delta^{ab}D(x-y)$.
Using the functional integral   approach, we introduce
bilocal field integration variables ${\cal B}^{\theta}(x,y)$ and
integrate out the quarks and gluons fields to change dynamical freedoms of
fundamental  QCD to hadronic variables. This method has
an obvious advantage that the effective hadron field
interactions retain knowledge of their subhadronic origin.
The resulting function $Z_{\rm GCM}$ takes the following form:
\begin{equation}
Z_{\rm GCM}=\int{\cal
D}{\cal B}e^{-S[{\cal B}]} ,
\end{equation}
 where the action is given by
\begin{equation}
 S[\bar{\eta},\eta,{\cal B}]=-{\rm TrLn}[G^{-1}]+\int d^{4}xd^{4}y
  \frac{{\cal B}^{\theta}(x,y){\cal B}^{\theta}(y,x)}{2g^2D(x-y)}
  +\int d^{4}xd^{4}y \bar{\eta}(x) G(x,y;[{\cal B}^{\theta}])\eta(y)
  ,
\end{equation}
and $G^{-1}$ is defined as
\begin{equation}
 G^{-1}(x,y)=( \gamma\cdot\partial+\psi(x) ) \delta(x-y)+\Lambda^{\theta}{\cal B}^{\theta}(x,y).
\end{equation}
Here the quantity $\Lambda^{\theta}$ arises from the Fierz reordering of the
current-current interaction term in Eq.(4) and represents
the direct product of Dirac, flavor SU(3) and color matrices.

The vacuum configurations are defined by minimizing the bilocal action:
$\left. \frac{\delta S[{\cal B}]}{\delta {\cal B}}
 \right |_{\bar{\eta} , \eta=0}=0,$
following this mean field approximation, we obtain
\begin{equation}
  {\cal B}^{\theta}_0(x-y)=g^2D(x-y){\rm tr}[\Lambda^{\theta}G_0(x-y)].
\end{equation}
These configurations provide self-energy dressing of the quarks through the
definition $\Sigma(p)\equiv\Lambda^{\theta}{\cal B}^{\theta}_0(p)=i \gamma \cdot p[A(p^2)-1]
+B(p^2)$. In fact, Eq. (8) is the rain-bow Schwinger-Dyson equation.
In this Letter, the Schwinger-Dyson functions $A(p^2)$ and
$B(p^2)$ are approximated by the results of ILM through equating the two point Green functions of the corresponding
quark in the two nonperturbative
methods.$^{\cite{Zong}}$

With the  GCM generation function Eq.(5), it is straightforward
to calculate the vacuum expectation value of any quark operator of the
form $O_{n}\equiv(\bar{q}_{i_{1}}\Lambda^{(1)}_{i_{1}j_{1}}q_{j_{1}})
(\bar{q}_{i_{2}}\Lambda^{(2)}_{i_{2}j_{2}}q_{j_{2}})\cdots
(\bar{q}_{i_{n}}\Lambda^{(n)}_{i_{n}j_{n}}q_{j_{n}})$
in the mean field vacuum.
Here $\Lambda^{(i)}$ represents the operator  in
Dirac, flavor and color space.

Taking the appropriate number of derivatives with respect to
the external sources
$\eta_{i}$ and $\bar{\eta_{j}}$ of Eq.(5)
and putting $\eta_{i}=\bar{\eta}_{j}=0$ ,
we obtain
\begin{equation}
\langle 0|:O_{n}:|0\rangle=(-)^{n}\sum_{p}[(-)^{p}\Lambda^{(1)}_{i_{1}j_{1}}
\cdots\Lambda^{(n)}_{i_{n}j_{n}}(G_0)_{j_{1}i_{p(1)}}\cdots(G_0)_{j_{n}i_{p(n)}}],
\end{equation}
where $p$ represents the permutation of the $n$ indices. In fact, the
above equation can only be correct with the assumption of vacuum saturate
 or vacuum dominance.$^{\cite{Shifman}}$ Furthermore, in calculation,
 we should subtract the perturbative part of the  two point
 Green function for the quark,$^{\cite{Zong1}}$ which is here neglected due to
 the small current quark mass.

 The  quark condensate  $\langle0| : \bar q(0) q(0) :|0 \rangle$
  can be inferred from Eq.(9),
\begin{equation}
\langle 0|: \bar q(0) q(0) :|0 \rangle = -\frac{12}{16 \pi^2}
\int_{0}^{ \mu} d s \frac {B(s)}{sA^2(s)+B^2(s)}.
\end{equation}
In the instanton liquid model, $A(q)=1$. Here $\mu$ is taken to be 1
$GeV^2$, the scale dependence of the quark condensate is thoroughly
discussed in the previous paper.$^{\cite{Wangwan}}$
The quark condensate is calculated for later use in
the expressions for the Chiral coefficients.

 We  expand the bilocal fields about the saddle point,
   \begin{eqnarray}
{\cal B}^{\theta}(x,y)={\cal B}^{\theta}_{0}
(x-y)+\hat{{\cal B}}^{\theta}(x,y),
   \end{eqnarray}
and then examine  the effective interactions of the fluctuations,
 $\hat{\cal B }$. The bilocal fields contain information about
 internal excitions of the $q \bar{q}$  in addition to  center
 of mass motion which is associated with the usual local field variables.
Thus, they can be projected
onto a complete set of internal excitations $\Gamma^{\theta}_{n}$
with the remaining center of mass degree freedom represented by
the coefficients
$\phi^{\theta}_{n}(P) $. In fact, the coefficients $\phi^{\theta}_{n}(P)
$ can be identified with the meson fields.
 The bilocal fluctuations can  be written as
 \begin{equation}
 \hat{\cal{B}}^{\theta}(P,q)=\sum\limits_{n}
 \phi^{\theta}_{n}(P)\Gamma^{\theta}_{n}(P,q) .
 \end{equation}

The functions $\Gamma^{\theta}_{n}$ are in general eigenfunctions
of the free kinetic operator of the bilocal fields. At the mass
shell point, $P^2=-M^{2}_{n}$, they satisfy the homogeneous
Bethe-Salpeter equation in the ladder approximation for the given
quantum number $\theta$ and modes n.$^{\cite{Tandy}}$

To obtain the chiral coefficients, we take a digression to discuss
Goldstone theorem. Dynamical chiral symmetry breaking is associated
with the occurrence of a massless Goldstone mode which can be related
to the dynamical generated  scalar self-energy of the quark
propagator in the limit of vanishing  current quark  mass. By
analysing the corresponding zero momentum behaviour of the axial-vector
Ward identity in the chiral limit, we can obtain the Goldberger-Treiman
relation for the quark-pseudoscalar vertex and then reduce
the bound state function $\Gamma^{\theta}_{n}(P,q)$
from ladder Bethe-Salpeter equation for
the pseudo-scalar Goldstone mode with $P^2=0$
to the quark self-energy function.$^{\cite{Frank}}$

Our main goal here is to obtain the effective action for the Goldstone
mode, the higher mass fluctuations present in the bilocal fields are
neglected. In fact, the higher mass fluctuations should have some effects
if they have been integrated out. Based on the above discussion,
we can write the full bilocal field of quark fields
in the following form$^{\cite{Frank}}$:
\begin{equation}
\Lambda^{\theta} {\cal B}^{\theta}=\Sigma(x-y)+B(x-y)
[U_{5}(\frac{x+y}{2})-1],
\end{equation}
where $U_{5}(x)=P_{R}U(x)+P_{L}U^{\dag} (x)$ with $P_{R,L}$ the
standard right-left projection operators. For the $SU(3)$ flavor
case under consideration here the chiral field $U$ is defined as
$U \equiv e^{i \lambda^{a} \phi^{a}/f_{\pi}}$ .

Next, we follow Ref. \cite{Frank} to calculate
the chiral coefficients and we consider only the real contribution
to the effective action.
The pioneer work for obtaining the chiral hadronic phenomenon through
global color model was given by Roberts et al.$^{\cite{CDRoberts}}$
The restriction of the fluctuations to the Goldstone modes with
$UU^{\dag}=1$  entails that the second term of the action in
Eq.(6) is independent of the filed $U$ and can be neglected.
Thus, the real contribution to the action can be written as
\begin{equation}
{\cal S} \equiv Re[S]=-\frac{1}{2} TrLn(G^{-1}[G^{-1}]^{\dag}),
\end{equation}
where
\begin{equation}
G^{-1}=\gamma \cdot
\partial_{x}A(x-y)+\psi(\frac{x+y}{2})\delta(x-y)+B(x-y)U_{5}(\frac{x+y}{2})
.
\end{equation}
Here $A(x-y) =\delta(x-y)$ in the ILM.

By expanding the logarithm and dropping the irrelevant constant terms,
we obtain the chiral Lagrangian to the fourth order (in Euclidean space):
\begin{eqnarray}
{\cal S}=&& \int d^4 x \{ \frac{f^2_{\pi}}{4}tr [
\partial_{\mu}U \partial_{\mu}U^{\dag} ]   -\frac{f^2_{\pi}}{4}
[ U \chi^{\dag}+\chi U^{\dag} ] \nonumber \\
&&-L_{1}(tr[\partial_{\mu}U \partial_{\mu}U^{\dag}])^2
-L_{2}tr[\partial_{\mu}U \partial_{\nu}U^{\dag}] tr[\partial_{\mu}U
\partial_{\nu}U^{\dag}] \nonumber \\
&&-L_{3}tr[\partial_{\mu}U \partial_{\mu}U^{\dag}\partial_{\nu}U
\partial_{\nu}U^{\dag}]
+L_{5}tr[\partial_{\mu}U \partial_{\mu}U^{\dag}(U \chi^{\dag}+\chi
U^{\dag})] \nonumber \\
&&-L_{8}tr[\chi U^{\dag}\chi U^{\dag}  +U \chi^{\dag}U \chi^{\dag}]
\},
\end{eqnarray}
where $\chi(x)=-2\langle \bar{q}q \rangle \psi(x)/f^{2}_{\pi}$ and the
remaining trace is over flavor. The coefficients  $L_{1}$,
$L_{2}$ $ (L_{2}=2L_{1})$, $L_{3}$, $L_{5}$, $L_{8}$ and $\pi$ decay
constant $f_{\pi}$ are functions
of the scalar part of quark self energy and its derives, as the
expressions are lengthy, one can see Ref.\cite{Frank} for details.

 In performing the gradient expansion it is
important to keep the $x$-dependence of the external field $\psi(x)$.
After carrying out the gradient expansion to the fourth order, we
employ the equation of motion which is obtained at the second order
and depends on the external field $\psi(x)$. Finally
we can identify $\psi(x)$ with the current quark mass matrix.
Failure to keep the $x$-dependence of $\psi$ to the very end
violates chiral invariance and will render some unphysical results for
 the low-energy coefficients.

From the above expressions, we can see that the coefficients $L_{4}$,
$L_{6}$, $L_{7}$  vanish, which is consistent with the large $N_{c}$ QCD
result. Some modifications  may be obtained if we have
integrated out higher mass states.

Instantons, as the solutions of the classical Yang-Mills equation of
motion, give many successful descriptions of strong interaction
phenomena, such as $U(1)_{A}$ problem, dynamical chiral symmetry
breaking, tunneling the $\theta$ vacuum and so on.
 After the great achievement of 't Hooft's
computing the functional determinant in the one-instanton
background, a tempt was made to understand quantitatively the
nonperturbative QCD vacuum configurations in terms of
instanton ensembles.$^{\cite{Hooft,Callen}}$
However, the notorious infrared problem due to large size
instanton made many researchers no longer believe in the dilute instanton gas
 model. To circumvent the disaster, one may think the overcome that
the vacuum gets more filled with instantons of
increasing size for larger distances, at some scale there might be some repulsive interactions
  to stabilize the medium while  the semiclassical treatment is still
  possible and the instantons are not much deformed through their
  interactions, thus form a dilute instanton liquid ensemble.
In Ref.\cite{Ilgenfritz,Diakonov1}, the authors  used
variational techniques and mean field approximation to deal with the
statistical mechanics of the instanton liquid while  stabilizing it by
phenomenological hard core and repulsive interactions derived from
ansatz of gauge field.  Subsequently, the
propagation of light quarks in the dilute instanton liquid ensemble was
formulated quantitatively and refined  due to the treatment of the
nonzero mode propagation.$^{\cite{Diakonov2,Pobylista}}$ Although it does not give rise to a long
range confining force between quarks, the instanton vacuum has been shown
to provide a good phenomenological description of many
hadronic properties.$^{\cite{Schafer,Hutter}}$
Furthermore, lattice calculations support that the vacuum gluon
 configurations can be
approximated by the dilute instanton liquid.$^{\cite{Negele}}$  Its shortcoming will be that
 the vector self-energy function of quark propagator
in the instanton ensembles $A(q^2)=1$ which is not  consistent with the corresponding
one in lattice QCD and other potential models.$^{\cite{Skullerud,Kisslinger}}$
 In addition, the
$q^2 \gg 1 GeV^2$ behaviour of the scalar self energy function $B(q^2)$ of
quark propagator in the instanton medium falls as $1/p^6$ which disagrees with QCD.
Thus, it is interesting to substitute the quark propagator which is dressed through
propagating in the dilute instanton medium into the expressions for
the $\pi$ decay constant and chiral coefficients. Our numerical
results fit experimental values well.
The quark propagator in the ILM can be written as
\begin{equation}
G(p)=[i\gamma \cdot p+M(p)]^{-1}+O(N/VN_{c}),
\end{equation}
where
\begin{eqnarray}
M(p)=\sqrt{\frac{N}{2VN_{c}}} \frac{p^2\varphi^{2}(p)}
{\sqrt{\int \frac{d^4 k}{(2\pi)^{4}}k^2 \varphi^{4}(k)}},
\nonumber \\
\varphi(k)=\pi \rho^2 \frac{\partial}
{\partial z}(I_{0}(z)K_{0}(z)-I_{1}(z)K_{1}(z))|_{z=k \rho/2},
\end{eqnarray}
where $I$ and $K$ are the modified Bessel functions, $N/V \approx (200 MeV)^{-4}$
and $\ \rho \approx (600MeV)^{-1}$.

In conlusion, we have provided a systematic way to combine
the GCM with the ILM.$^{\cite{Zong}}$
 Then based on the nonperturbative quark propagator derived in the ILM,
 we calculate chiral coefficients which are expressed in terms of the
 scalar part of
 quark self energy and its derives.

  Our results are compatible with the
experimental data. Our results are $f_{\pi}=86 MeV$, $\langle \bar{q}
q \rangle=-(217MeV)^3$, $L_{1}=0.8 \times 10^{-3}$, $L_{2}=1.6 \times 10^{-3}$,
$L_{3}=-3.9\times 10^{-3}$, $L_{5}=2.0\times 10^{-3}$,
 $L_{8}=1.0\times 10^{-3}$. The experimental values are
 $L_{1}=0.9\pm 0.3 \times 10^{-3}$, $L_{2}=1.7 \pm 0.7 \times 10^{-3}$,
$L_{3}=-4.4 \pm 2.5 \times 10^{-3}$, $L_{5}=2.2 \pm 0.5 \times 10^{-3}$,
 $L_{8}=1.1 \pm 0.3\times 10^{-3}$

Following the GCM, we truncate the QCD Lagragian to include
only two point gluon Green function which is approximated by the
dilute instanton liquid ensemble and describe the mesons as fluctuations
about the saddle point of the effective action. The chiral
coefficients obtained in this way are consistent with the large
$N_{c}$ QCD, and embody their underlying theory through the
dependence on the quark self-energy and its derives. Some
modification will be made if we have integrated out the high mass
fluctuations instead of  neglecting them.

To obtain the nonperturbative quark propagator, we have to resolve the
rain-bow Schwinger-Dyson equation. However, the form of
 the gluon propagator $g^2G(p)$ in the infrared region is unknown,
 one often uses model forms as input in the previous studies
  of the rainbow Schwinger-Dyson equation.$^{\cite{Cahill,Tandy,Roberts}}$
  Obviously,  this dressing comprises the notation of constituent quarks by
 providing a mass $m(p^2)=B(p^2)/A(p^2)$ with $(m(0)=418MeV)$ in the ILM, which is corresponding to
 dynamical symmetry breaking. Here we use the dilute
 instanton liquid to approximate the QCD vacuum configurations,
 the quark propagator is dressed through propagating in the dilute
 instanton liquid ensemble and contains valuable nonperturbative
 information which manifest
 itself in the coefficients. Although ILM has shortcoming as
 mentioned above, it does give satisfactory chiral
 coefficients and can be extended to other systems. Furthermore,
 the lattice calculations support that the dilute instanton liquid ensemble is a
 good approximation of vacuum configurations.

\end{document}